\documentstyle[preprint,aps,prl,psfig]{revtex}
\tightenlines
\begin{document}
\newcommand{\sinf}{\raisebox{-.7ex}{$\stackrel{<}{\sim}$}}
\newcommand{\ssup}{\raisebox{-.7ex}{$\stackrel{>}{\sim}$}}

\ifx\undefined\psfig\def\psfig#1{    }\else\fi
\ifpreprintsty\else
\twocolumn[\hsize\textwidth
\columnwidth\hsize\csname@twocolumnfalse\endcsname       \fi    \draft
\preprint{  }  \title  {Spin Coulomb Drag}
\author  {Irene D'Amico and Giovanni Vignale}
   \address{Department  of Physics,   University   of
Missouri, Columbia, Missouri 65211} \date{\today} \maketitle

\begin{abstract}
We introduce a distinctive feature of spin-polarized transport,
the Spin Coulomb Drag: there is an {\it intrinsic} source of friction for
spin currents  due to the Coulomb interaction between spin ``up'' and spin
``down'' electrons. We calculate the associated ``spin transrestistivity''
in a generalized random phase approximation  and  discuss its dependence on
temperature, frequency,  and electron density. We show that, in an appropriate
range of parameters,  such resistivity is measurable and propose an
experiment to measure it.

\end{abstract}
 \ifpreprintsty\else\vskip1pc]\fi \narrowtext
Interest in spin-polarized transport has been growing dramatically in the last
few years, spurred by the hope of realizing practical spin-electronic
devices in
a not too distant future  \cite{spintr}. In particular it has been shown
that spin coherence can be maintained over large distances $\delta_s
\stackrel{>}{
 \sim} 100 \mu m  $
and for long times $T_2 \sim 10^{-9}-10^{-8}s$ both in metals
\cite{Silsbee} and in
semiconductors \cite{Awscha}.

In this paper
we  introduce  a distinctive feature of  spin-polarized transport:
in a conductor, due to the Coulomb interaction, there is an {\it intrinsic}
mechanism for friction between electrons of different spin,  the {\it
``Spin
Coulomb Drag''} (SCD). For simplicity  we shall restrict our discussion to the
case in which the spin state of each electron can be classified as
``up'' or  ``down'' relative to the $z$ axis. In the absence of
impurities the total momentum ${\bf P }=\sum_i {\bf p
}_i$, where ${\bf p
}_i$ is the momentum of the  $i-th$ electron,  is a conserved quantity. On
the contrary, the ``up'' and the ``down'' components of the total momentum
${\bf
P}_{\uparrow}=\sum_i {\bf p }_{i\uparrow} {1 + \hat \sigma_{zi} \over 2}$,
and ${\bf P}_{\downarrow}=\sum_i{\bf p
}_{i\downarrow}{1 - \hat \sigma_{zi} \over 2}$, where $\hat \sigma_{zi}$
is the the Pauli matrix for the $z$ component of the $i-th$ electron's spin,
 are  {\it not separately conserved} even {\it in the absence of impurities}:
Coulomb scattering can transfer momentum between
spin up and spin down electrons   thereby
effectively introducing a ``friction"  for relative motion of the
two spin components.
If, for example,   one of the two spin
components is set into motion relative to the other, it will tend to drag
the latter in the same direction.  Or, if a finite spin current is
set up through the application of an external field, then the Coulomb
interaction
will tend to equalize the net momenta of the two spin components,  causing the
difference $\langle P_\uparrow \rangle  - \langle P_\downarrow \rangle$  to
decay
to zero  when  the external field is turned off.

 The most dramatic manifestation of the SCD is the appearance
of a finite {\it trans-resistivity} defined as the
ratio of the gradient of the spin-down electro-chemical
potential to the spin-up current density when the spin-down current is zero.
This is completely  analogous to the trans-resistivity measured in Coulomb drag
experiments  with electrons in  two separate
layers \cite{dragold,dragexpts,Rojo}, but in this case what makes the two
electron populations
distinguishable  is not a physical separation but the different spin. In
SCD the
non conservation of the spin, caused mainly by the spin-orbit
interaction, represents a ``leakage'' mechanism  analogous to
the interlayer tunneling in the usual Coulomb drag.

First of all let  us describe the SCD from a phenomenological point of
view.
Let ${\bf E}_\uparrow(t)$ and  ${\bf E}_\downarrow(t)$ be uniform effective
electric fields that couple to spin $\uparrow$ and spin $\downarrow$  electrons
respectively.  These fields are sums of the ordinary electrostatic field
plus the gradient of the local chemical potential, which can be
spin-dependent.
We  restrict ourselves to the linear response
regime, assume weak electron-electron and electron-impurity
scattering, and ignore spin-flipping processes altogether.   If ${\bf
v}_\sigma$
is the velocity of the center of mass of electrons of spin $\sigma$, and
$N_\sigma$ the number of such electrons, the phenomenological equation of
motion has the form
\begin{equation}
mN_{\sigma} \dot{{\bf v}}_\sigma = -eN_{\sigma} {\bf E}_\sigma +
{\bf F}_{\sigma \bar{\sigma}} -\frac{m}{\tau_D} N_{\sigma} {\bf v}_\sigma
\label{eqmot}
\end{equation}
where $\tau_D$ is the Drude scattering time and ${\bf F}_{
\sigma\bar{\sigma}}$ is the Coulomb force exerted by spin $\bar{\sigma}
(= - \sigma)$ electrons on spin $\sigma$ electrons.  By Newton's third
law ${\bf F}_{\sigma\bar{\sigma}}=- {\bf F}_{\bar{\sigma}\sigma}$, and by
Galilean invariance this force can only depend  on the relative velocity
of the two components.  Hence, for weak Coulomb coupling we  write
\begin{equation}
{\bf F}_{ \sigma\bar{\sigma}}=
- \gamma m N_\sigma  \frac{n_{\bar\sigma}}{n}({\bf
v}_\sigma-{\bf v}_{\bar{\sigma}}), \label{gamma1}
\end{equation}
where $n_\sigma$ is the number density of electrons of spin  $\sigma$ and
$n=n_\uparrow + n_\downarrow$ is the total density.
Eq.~(\ref{gamma1}) defines the {\it spin drag coefficient} $\gamma$.
Fourier transforming Eq.~(\ref{eqmot}) with respect to time, and making use
of the relationship  ${\bf j}_\sigma(\omega)=
-e n_\sigma{\bf v}_\sigma(\omega)$ between current density and velocity,
we obtain
\begin{eqnarray} \label{eqmotf2} i\omega{\bf
j}_\sigma(\omega) = &-& \frac{n_\sigma e^2}{m}{\bf E}_\sigma (\omega)
+\left ( \frac{n_{\bar\sigma}}{n} \gamma + \frac {1} {\tau_D} \right ){\bf
j}_\sigma (\omega) \nonumber \\&-&\frac{n_\sigma}{n}\gamma{\bf
j}_{\bar{\sigma}} (\omega).
\end{eqnarray}
The resistivity matrix $\rho_{\sigma,\sigma'}$ is defined as the coefficient of
proportionality between the electric field and the current: ${\bf E}_\sigma =
\sum_{\sigma'}\rho_{\sigma,\sigma'} {\bf j}_{\sigma'}$.  A quick comparison
between this definition and
Eq.~(\ref{eqmotf2}) shows that $\gamma$ is directly proportional to the spin
trans-resistivity
\begin{equation} \gamma = -{ ne^2\over m } \rho_{\uparrow
\downarrow}.
\label{gamma}
\end{equation}

Let us now proceed to the microscopic calculation of the spin
trans-resistivity.
We start from the Kubo formula \cite{Kubo} for the uniform  conductivity matrix
\begin{equation}
\label{Kuboconductivity}
\sigma_{\sigma, \sigma'} (\omega)= -{1 \over i \omega} {e^2 \over m} [n_\sigma
\delta_{\sigma, \sigma'} + {\langle \langle {\bf P}_\sigma ; {\bf P}_{\sigma'}
\rangle \rangle_\omega \over m} ],
\end {equation}
where  $\langle \langle A;B \rangle \rangle_\omega$ represents, as usual
\cite{Kubo}, the retarded  response function for the expectation value of $A$
under the action of a field that couples linearly to $B$.  The resistivity
matrix is the inverse of the conductivity matrix.  In the limit that both the
electron-impurity and the electron-electron scattering are weak the
$P_\sigma$'s are almost constants of the motion and therefore
$\langle \langle {\bf P}_\sigma ; {\bf P}_{\sigma'} \rangle \rangle_\omega \to
0$. This means that the second term in the square
bracket of Eq.~(\ref{Kuboconductivity}) can be treated as a small correction
to the first \cite{Goetze}.  Inverting Eq.~(\ref{Kuboconductivity}) to first
order in  $\langle \langle {\bf P}_\sigma ; {\bf P}_{\sigma'} \rangle
\rangle_\omega$ and selecting the $\uparrow \downarrow$ matrix element we
obtain
\begin{equation}
\label{transresistance1}
\rho_{\uparrow \downarrow} (\omega) = - {i \omega \over e^2}
 {\langle \langle {\bf P}_\uparrow ; {\bf P}_\downarrow
\rangle \rangle_\omega \over n_\uparrow n_\downarrow}.
\end {equation}
It is convenient to recast this equation in a form that emphasizes the
importance of the non conservation of $P_\uparrow$ and $P_\downarrow$.
To this end we make use twice of the general equation of motion
\begin{equation}
\langle \langle A;B \rangle
\rangle_\omega=\frac{1}{ \omega}(\langle[A,B]\rangle +i\langle
\langle \dot{A};B \rangle \rangle _\omega),
\end{equation}
where  $\dot{A} \equiv i [A,H]$ is the time derivative of the operator $A$, and
$\langle .. \rangle$ denotes the  thermal average.
Thus, Eq.~(\ref{transresistance1}) can be rewritten as
\begin {equation}
\label{transresistance2}
\rho_{\uparrow \downarrow} (\omega) =  -{i \over e^2 n_\uparrow
n_\downarrow} {\langle \langle \dot{{\bf P}}_\uparrow ; \dot{{\bf
P}}_\downarrow
\rangle \rangle_\omega + i\langle[\dot{{\bf P}}_\uparrow,{\bf
P}_\downarrow] \rangle  \over  \omega}.
\end {equation}
The commutator term  controls the high frequency
behavior of  $\rho_{\uparrow \downarrow} (\omega)$ and can be   expressed
 in terms of ground-state properties \cite{Goodman}.  This term however gives a
purely imaginary contribution to the trans-resistivity.  Our present
interest is
in the real part of the trans-resistivity, which is controlled by the imaginary
part of the force-force response function.

 The force operator is given by
\begin{equation}
\label{force}
\dot{{\bf P}_\sigma}= -\frac{i}{V}\sum_{\bf q}{\bf q}v_q
\rho_{{\bf q}\bar{\sigma}} \rho_{-{\bf q}\sigma} -\frac{i}{V}\sum_{\bf q}{\bf
q}v^{e-i}_q\rho^i_{{\bf q}}\rho_{-{\bf q} {\sigma}},
\end{equation}
where $v_{ q} = 4 \pi e^2/q^2$ is the Fourier transform of the Coulomb
interaction, $v^{e-i}_q$ is the Fourier transform of the
electron-impurity interaction, $\rho_{{\bf q}\sigma}$ is the electronic spin
density fluctuation operator, $\rho^i_{{\bf q}}$ is the Fourier transform
of the
impurity density (a number),  and $V$ is the volume of the system.

In calculating the force-force response function special attention must be paid
to the contributions of the electron-impurity interaction.  In the theory
of the
ordinary Coulomb drag \cite{recenttheory} such contributions are zero
on the average because the electrons in the two layers interact with two
{\it different} sets of impurities, which are uncorrelated to each
other.  In the present case, however, electrons of opposite spin
interact with the {\it same} set of impurities, so that electron-impurity
terms, generated from the substitution of Eq.~(\ref{force}) into
Eq.~(\ref{transresistance2}),  do not vanish upon disorder averaging.
Happily, it turns out that these terms cancel out exactly at low
frequency ($\omega <<E_F$) and to leading order in the
electron-electron and electron-impurity interactions \cite{footnote1}.
Thus, the real part of the spin trans-resistivity takes the form
\begin{eqnarray} \label{realtransresistance}
Re \rho_{\uparrow \downarrow} (\omega)&=&-\frac{1}{n_\uparrow
n_{\downarrow}e^2\omega V^2}\sum_{\bf q q'}\frac{\bf q \cdot q'}{3} v_{q}v_{
q'}\cdot \nonumber \\ &~&Im  \langle \langle \rho_{-{\bf q}\uparrow}\rho_{{\bf
q}\downarrow};\rho_{{\bf
q'}\uparrow}\rho_{-{\bf q'}\downarrow} \rangle \rangle_\omega,
\label{gamma2}
\end{eqnarray}
where we have made use of the isotropy of the electron gas.

We have calculated the four point response function \\
$\chi_{4\rho}({\bf q},{\bf q'},\omega) \equiv \langle \langle \rho_{-{\bf
q}\uparrow}\rho_{{\bf q}\downarrow};\rho_{{\bf
q'}\uparrow}\rho_{-{\bf q'}\downarrow} \rangle \rangle_\omega$ at finite
temperature in a generalized Random Phase
Approximation (RPA). The selected diagrams are shown  in Fig. \ref{fig1}.
Because of its
infinite range, the Coulomb interaction must be treated to infinite order, even
when weak.  The sum of the RPA diagrams has
been evaluated by standard methods \cite{Mahan} with the following result:
\begin{eqnarray}
&~& Re \rho_{\uparrow\downarrow}(\omega,T) =
-\frac{1}{n_\uparrow n_\downarrow V e^2}\sum_{\bf q }\frac{q^2}{3}
v_{q}^2\cdot\frac{(e^{-\beta\omega}-1)}{\omega}
\\ &~& 
\int_{-\infty}^\infty \frac{d\omega '}{\pi}
\frac{[\chi''_{\uparrow\uparrow}(q,\omega ')
\chi''_{\downarrow\downarrow}(q,\omega-\omega ')
- \chi''_{\uparrow\downarrow}(q,\omega ')
\chi''_{\downarrow\uparrow}(q,\omega-\omega ')]}{(e^{-\beta\omega
'}-1)(e^{-\beta(\omega-\omega ')}-1)}. 
\label{rho}
\end{eqnarray}

Here $\beta=1/k_B T$, with $k_B$ the Boltzmann constant, $\chi''_{\sigma
\sigma'}(q,\omega)$ is the imaginary part of the RPA
spin-resolved density-density response function, which is related to the
noninteracting response function $\chi_{0\sigma} (q,\omega)$ as
follows
\begin{equation}
\label {chirpa}
[\chi^{-1}(q, \omega)]_{\sigma \sigma'} = [\chi_{0 \sigma}]^{-1}
(q,\omega) \delta_{\sigma \sigma'} - v_{q}.
\end{equation}

It is possible to show by simple but tedious algebraic calculations
that this expression for the spin trans-resistivity
$\rho_{\uparrow\downarrow}(\omega,T)$ reduces, in the case of finite
temperature
and $\omega=0$, to the well know result of memory function and diagrammatic
theories for the Coulomb drag \cite{recenttheory}, \cite{Forster}.
Furthermore, for $T=0$ and $\omega\ne 0$, the RPA
is equivalent to the decoupling approximation for the four-point
response function used in \cite{Nifosi} to calculate the dynamical
exchange-correlation kernel.
Thus our calculation demonstrates that those two approximations, quite
different at a first sight, are  simply  RPAs performed in different limits.

Let us
focus on the low-temperature and low-frequency regime  $k_BT << E_F$ and
$\omega <<E_F$, with $E_F$
the Fermi energy. In this regime the imaginary part of the  density-density
response functions  $\chi''_{\sigma
\sigma'}(q,\omega)$ is a linear function of $\omega$. 
In the limit of vanishing impurity concentration $\chi_{0 \sigma}(q,
\omega)$ is simply the Lindhard function, whose imaginary part, at low
frequency, is given by
$\chi_{0\sigma}''({\bf q},\omega\to 0)=-(m^2/4\pi)(\omega/q)$ and whose
real
part can be approximated by its value at $\omega=0$.
Making use of this limiting form, the calculation of $\rho_{\uparrow
\downarrow}$ can be carried in an essentially analytical fashion. The
  result is
\begin{eqnarray}
Re\rho_{\uparrow\downarrow}(\omega,T) &=&{\hbar a \over
e^2}{4\pi^2 (k_BT)^2 + \omega^2 \over 6 (Ry)^2}\cdot \nonumber
\\ &~&{1\over 24\pi^3
\bar{n}_\downarrow\bar{n}_\uparrow}
\int_{-\infty}^{\infty}{d\bar{q}\over \bar{q}^2} {1\over |\epsilon
(\bar{q}/a,0)|^2}
\label{rholim}
\end{eqnarray}
where $a\equiv \hbar^2/me^2$ is the effective Bohr radius, $Ry=e^2/2a$ is the
effective Rydberg, $\bar{q}\equiv qa$, $\bar{n}_\sigma\equiv n_\sigma a^3$
and
 $\epsilon (q,\omega) = 1-v_q\chi_{0\uparrow}({\bf q},
\omega)-v_q\chi_{0\downarrow}
({\bf q}, \omega)$ is the RPA
dielectric function.
Eq.~(\ref{rholim}) shows that, in the absence of impurities,
$\rho_{\uparrow\downarrow}(\omega,T)$ is proportional to $\omega^2$
for $k_BT<<\omega$ and to $T^2$ for $\omega<<k_BT$.

Modifications in the form of $\chi_{0 \sigma}(q, \omega)$ due to the presence
of impurities can be taken into account through Mermin's approximation scheme
\cite{Mermin}. These modifications amount to replacing $\omega/q v_F$ by
$\omega /Dq^2$  ($D = v_F^2 \tau/3$ being the diffusion constant) for $\omega <
1 /\tau$ and $q<1/v_F \tau$ where $v_F$ is the Fermi velocity and $\tau$ is the
electron-impurity mean scattering time.  The $\omega$
and $T$ dependencies of Eq.~(\ref{rholim})  are not affected.

Writing  explicitly in Eq.~(\ref{rholim}) the dependence over $r_{s\sigma}$
(where $r_{s\sigma}=(4\pi n_\sigma a^3/3)^{-1/3}$ is the usual electron gas
parameter)
 one can also see that  $
\rho_{\uparrow\downarrow}(\omega,T) \sim
r_{s\uparrow}^3r_{s\downarrow}^3$, so that $
\rho_{\uparrow\downarrow}  $ will strongly increase  with
decreasing electron density. In Fig. \ref{fig2} we plot $
\rho_{\uparrow\downarrow}(\omega=0,T)$ for $n_\uparrow=n_\downarrow$, at
metallic
densities (roughly $2<r_s<6$) in the temperature range  $10 K <T<70 K$. It can
be seen that for temperatures of the order of $40-60K$ (at which for example
experiments on spin relaxation time using spin polarized currents have been
performed \cite{Silsbee}), the spin trans-resistivity is appreciable ($
\rho_{\uparrow\downarrow}(\omega=0,T)\stackrel{>}{
 \sim} 0.01\mu\Omega cm$).

In the remaining part of this paper, we  describe an experiment aimed at
detecting the effect of the Spin Coulomb Drag and measuring the
spin trans-resistivity.   The setup is shown in Fig.3: a paramagnetic metal
film
of thickness $L$ is sandwiched between two ferromagnets polarized in the same
 direction. A battery is connected to the ferromagnets inducing a
{\it spin-polarized} current \cite{Silsbee} from the first ferromagnet
(``injector'')
through the paramagnet and  toward the second ferromagnet (``receiver''). The
injector and receiver are chosen to be {\it semi-metals}, i.e, they have only
electron states of spin $\uparrow$ at the Fermi level (see Fig.\ref{fig3}).  It
follows that the injected current ${\bf j}_\uparrow$ is carried {\it only} by
spin $\uparrow$ electrons. If we choose  $L<<\delta_s$,  where $ \delta_s$ the
spin relaxation length, we can safely neglect spin-flipping processes and the
polarized current entering the paramagnet will not relax before reaching  the
receiver. Spin relaxation lengths are relatively large in some materials ($
\delta_s\approx 100\mu m$ in Al  \cite{Silsbee}), so the condition
$L<<\delta_s$
is not particularly restrictive. Due to the SCD, the injected
${\bf j}_\uparrow$ will drag spin $\downarrow$ electrons toward the junction
with the receiver. But, since there is
no conduction band available in the receiver for spin $\downarrow$ electrons
the circuit will behave as  an {\it open circuit} for  spin
$\downarrow$ electrons, i.e., ${\bf j}_\downarrow=0$.  The vanishing of ${\bf
j}_\downarrow$ is an indication that the Coulomb drag force is exactly balanced
by the gradient of the electro-chemical potential for spin down
\begin{equation}
\label{forcebalance} -e{\bf E}_\downarrow+m\gamma\frac{{\bf
j}_\uparrow}{n_\uparrow}=0. \end{equation}
where ${\bf E}_\downarrow=\nabla \mu_\downarrow /e +{\bf E}$ is the sum of the
electrostatic field  ${\bf E}$  and the gradient of the chemical potential
$\mu_\downarrow$.  What Eq.~(\ref{forcebalance}) tells us
is that due to the SCD  there will be a measurable electro-chemical potential difference
 $eE_{\downarrow} l = e m \gamma j_\uparrow l/n_\uparrow$ for spin $\downarrow$
electrons between two points within the metal separated by a distance $l$ along
the direction of the current.

To measure this potential difference a second circuit including a
voltmeter of  very large  resistance is connected to the regions of the
paramagnet close to the junctions (See Fig. 3). Our purpose is to measure
$E_\downarrow$, so this second circuit
must be driven by the spin $\downarrow$  electro-chemical potential only.
In order to
accomplish this, we propose to use as contacts  two semi-metallic
ferromagnetic electrodes  (``detectors''),  similar to the injector and the
receiver, but polarized in the {\it opposite} direction. In this way, for the
same reasons explained before, the detection circuit will be ``open"  as far as
spin $\uparrow$ electrons are concerned,  and the current flowing in the
voltmeter will be exclusively driven by the electro-chemical potential
difference
of spin $\downarrow$ electrons.  The spin trans-resistivity will then be
given by
$\rho_{\uparrow\downarrow}= (\Delta V_D /I_\uparrow)(A/l)$, where $\Delta V_D$
is the voltage measured by the meter,  $A$ is the cross-section of the
paramagnetic metal, $l$ is the distance between the detectors, and $I_\uparrow$
the current flowing between injector and receiver.  As shown by our
calculations, we expect a resistivity of the order of $10^{-2}\mu\Omega cm$
that is proportional to $T^2$ for $k_B T>>\omega$.

In summary we have pointed out a novel effect in spin-polarized
transport,  the Spin Coulomb Drag,  and  have proposed an experiment to
observe it.  We hope that this paper will stimulate experimental work
aimed to the detection of this effect.

This research was supported by NSF Grant No. DMR-9706788.  We thank  Shufeng
Zhang for very valuable discussions.

%*******************
\newpage
\begin{figure}
  \caption{The two series of ``bubble'' diagrams for the four-point response
function $\chi_{4 \rho}$ in the RPA. The vertices represent spin-density
fluctuations
$\rho_{q \sigma}$ as labelled.}
  \label{fig1}
\end{figure}

\begin{figure}
  \caption{Density and temperature dependence of
$\rho_{\uparrow\downarrow}(0,T)$ in a paramagnetic metal.
Top and bottom lines correspond to  $T=70 K$ and $T=10K$ respectively.
Temperature is incremented in steps of $10 K$ starting from the bottom.}
  \label{fig2}
\end{figure}

\begin{figure}
  \caption{(a) Experimental setup to detect the SCD effect: the voltage
$\Delta V$
is applied between two parallel semi-metallic ferromagnets (injector (inj.) and
receiver (rec.)) that sandwich a paramagnet (P). The voltage $\Delta V_D$ is
detected using two ferromagnetic electrodes (d) similar to the injector and the
receiver, but polarized in the {\it opposite} direction.  (b) Schematic
bandstructure of injector, receiver, d and P. 
}
\label{fig3}
\end{figure}
%*******************
\newpage
\begin{figure}
  \psfig{figure=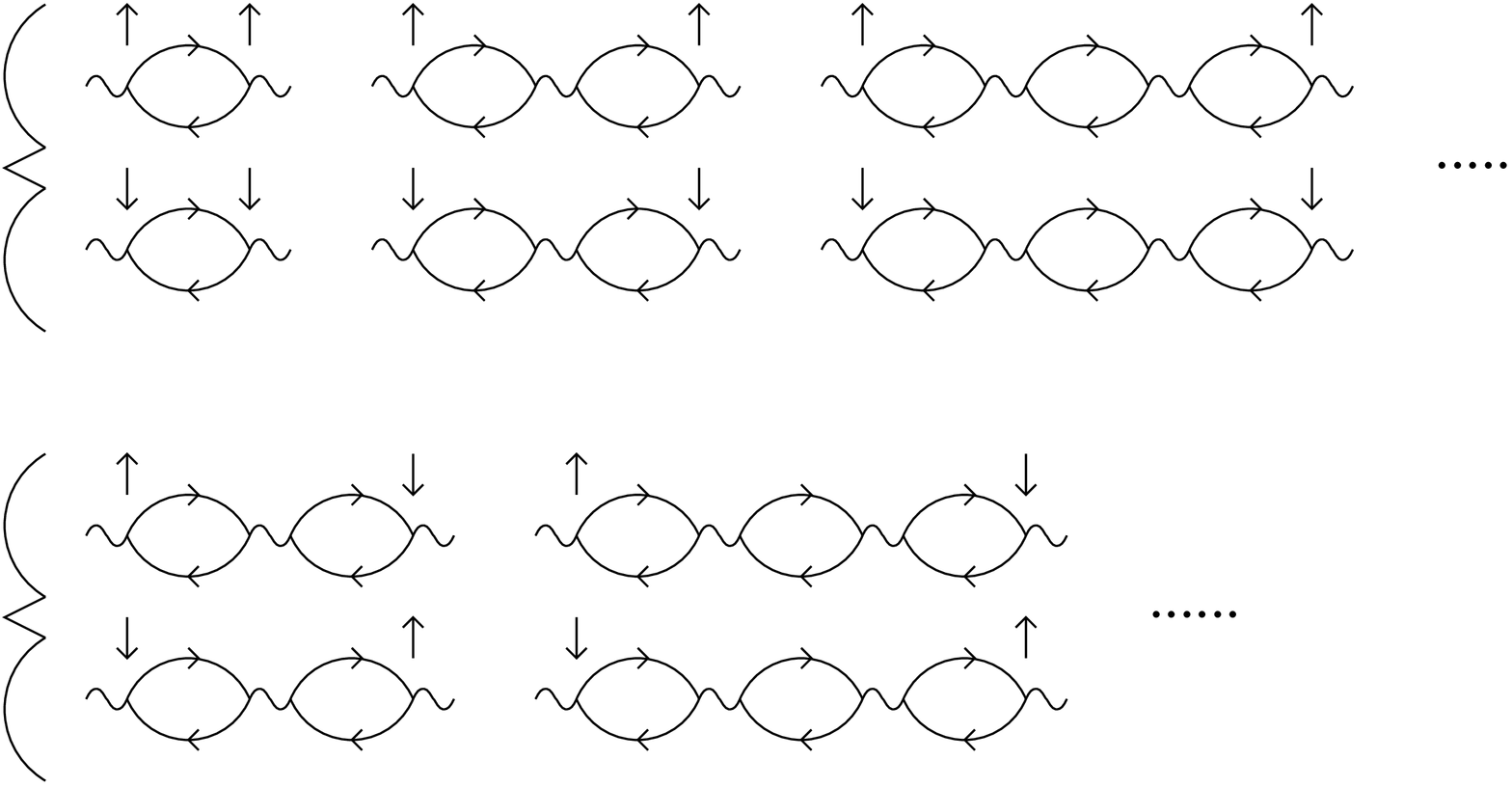,width=0.60\columnwidth,angle=-90}
\end{figure}
\Large{Fig. 1}

\newpage
\begin{figure}
  \psfig{figure=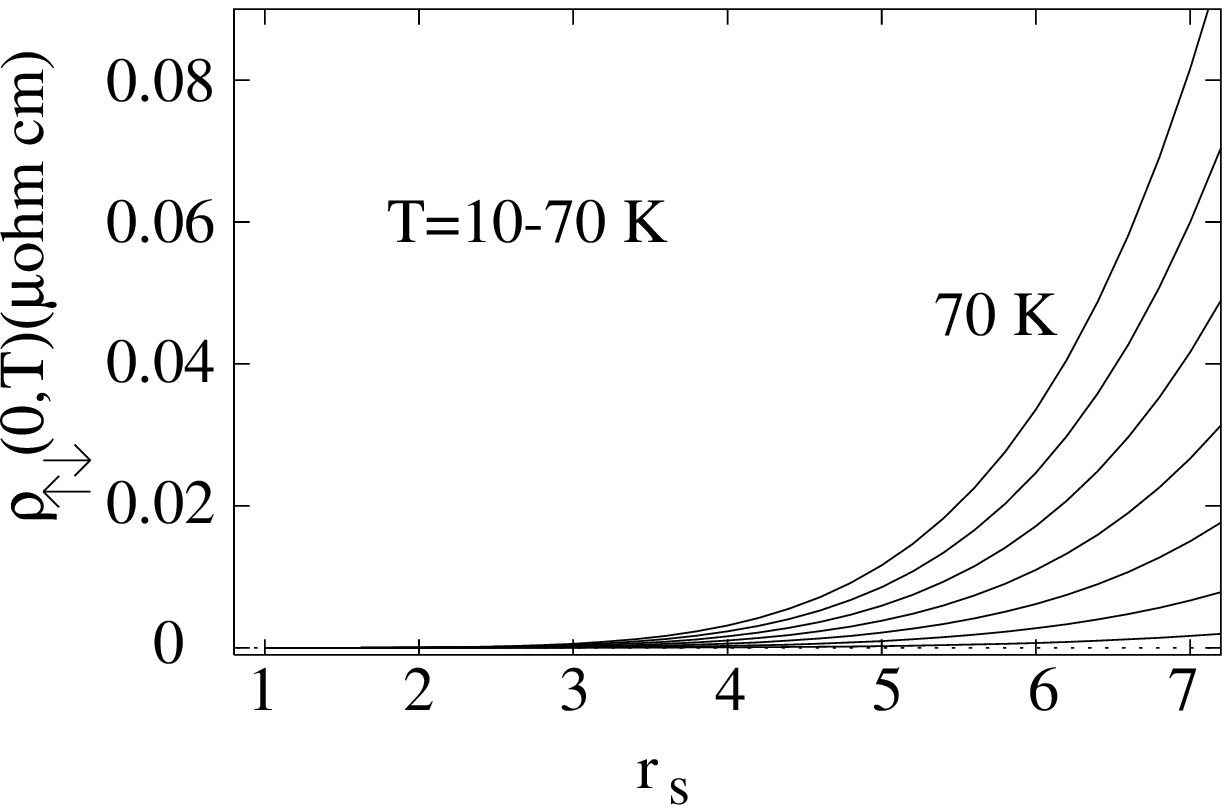,width=0.70\columnwidth}
\end{figure}
\Large{Fig. 2}

\newpage
\begin{figure}
\psfig{figure=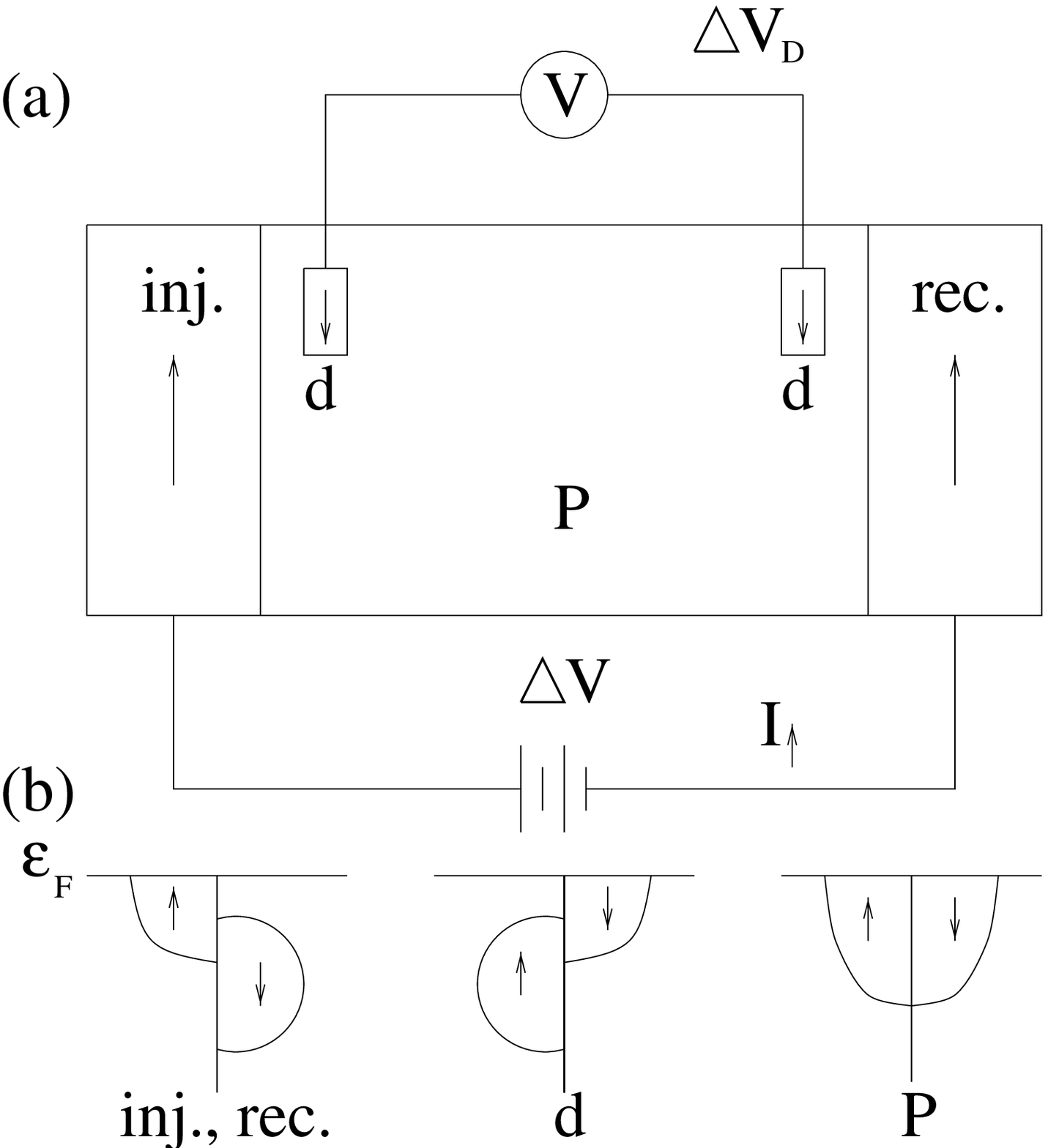,width=0.65\columnwidth}
\end{figure}
\Large{Fig. 3}

\end{document}